\documentclass[twocolumn,showpacs,preprintnumbers,amsath,assymb]{revtex4}
\usepackage{epstopdf}
\usepackage{graphicx}
\usepackage{dcolumn}
\usepackage{bm}

\def\gappeq{\mathrel{ \rlap{\raise.5ex\hbox{$>$}}
                      {\lower.5ex\hbox{$\sim$}} } }
\def\lappeq{\mathrel{ \rlap{\raise.5ex\hbox{$<$}}
                      {\lower.5ex\hbox{$\sim$}} } }

\begin{document}

\preprint{PRA}

\title{Bright solitary waves of atomic Bose-Einstein condensates under rotation}

\author{N.A. Jamaludin$^{1}$, N.G. Parker$^{1,2}$ and A.M. Martin$^1$}

\address{$^{1}$ School of Physics, University of Melbourne, Parkville,
Victoria 3010, Australia. \\$^{2}$ Department of Physics and
Astronomy, McMaster University, Hamilton, Ontario, L8S 4M1, Canada.}

\date{\today}

\begin{abstract}
We analyse the rotation of bright solitary waves formed of atomic
Bose-Einstein condensates with attractive atomic interactions. By
employing a variational technique and assuming an irrotational
quadrupolar flow field, we map out the variational solutions in the rotating frame. In particular, we show that rotation has a considerable stabilising effect on the system, significantly raising the critical threshold for collapse of the bright solitary waves.
\end{abstract}

 \maketitle

In recent years, bright solitary waves have been created
using ultra-cold atomic Bose-Einstein condensates (BECs)
\cite{strecker,khaykovich,cornish}. Under attractive atomic
interactions, these matter waves are self-trapped in the longitudinal
direction \cite{perez_gaussian,perez_garcia,carr,salasnich,parker}
and are closely analogous to the classic one-dimensional soliton
\cite{shabat}. However, these states must be confined in the
remaining directions by a waveguide and can
retain three-dimensional effects. The most lucid example is
the collapse instability: in 3D a homogeneous, untrapped BEC
with attractive interactions is unstable to collapse \cite{nozieres}
while the 1D limit is stable to collapse. The presence of external
trapping stabilises the BEC up to a critical atom number
(or interaction strength) before collapse is triggered, as
demonstrated experimentally \cite{bradley,roberts}. The collapse instability has limited bright solitary
wave (BSWs) experiments to only a few thousand atoms per BSW
\cite{strecker,khaykovich,cornish}.  The BSW solutions, and their critical points, have been studied
theoretically \cite{carr,parker,salasnich,perez_garcia}, with
variational approaches shown to give very good predictions.
The collisions of BSWs, which show intriguing behaviour and may have applications in interferometry, are also prone to collapse instabilities
\cite{parker,salasnich,brand,adhikari_NJP,BSW_collisions}. As such,
it is pertinent to consider approaches to suppress collapse in
attractive BECs. Periodic modulation of the interaction strength,
made possible by employing a Feshbach resonance, is predicted to
partially stabilise against collapse \cite{feshbach}, although when
the average interaction is attractive, collapse is inevitable
\cite{konotop}. The presence of a quantized vortex is also predicted
to raise the threshold for collapse \cite{Dalfovo,adhikari_NJP},
although the presence of a vortex in an attractive condensate under
harmonic trapping is not energetically stable \cite{Dalfovo,Wilkin}.

Due to the superfluid nature of BECs, it is intriguing to study
their response to rotation. The rotation of repulsive BECs has been
considered extensively both experimentally and theoretically (see
\cite{book} for a review). One method is to mechanically rotate the
BEC in an elliptical trap \cite{madison1,madison2,hodby} formed by
time-dependent laser or magnetic fields. At low rotation frequency
$\Omega$, the condensate remains irrotational and vortex-free.
According to a hydrodynamical model, the BEC can access a family of
rotating stationary solutions, characterised by a quadrupolar
irrotational flow pattern \cite{Recati,Sinha,Rapid,Castin,PRL,corro}
and confirmed experimentally \cite{madison2}. At a critical rotation
frequency, which coincides with when these irrotational solutions
become unstable \cite{Rapid,corro}, vortices are nucleated and form
a vortex lattice. In the context of attractive BECs, theoretical
work has shown that a centre-of-mass mode is favoured under rotation
rather than the occurence of vortices \cite{Wilkin,collin}. Under
harmonic trapping, this mode becomes excited when $\Omega$ exceeds
the trap frequency.

In this work we consider the rotation of bright solitary matter
waves in an elliptical trap about the longitudinal axis. We employ a
variational technique based on assuming an ansatz for the BSW
profile which incorporates a quadrupolar irrotational flow pattern.
By deriving the variational energy of the system, we obtain the BSW
solutions and analyse their response to rotation.  In particular, we
find that rotation of the BSW can significantly increase the
critical point for collapse.

We consider a BEC confined by an atomic ``waveguide" potential, under
rotation at frequency $\Omega$ about the longitudinal axis. In the
limit of zero temperature, the BEC can be described by a mean-field
``wavefunction" $\Psi({\bf r},t)$ which satisfies the
Gross-Pitaevskii equation (GPE) \cite{book},
\begin{eqnarray}
i \hbar \frac{\partial \Psi}{\partial t}=
\left[-\frac{\hbar^2}{2m}\nabla^2 + V({\bf r})+g|\Psi|^2 -\Omega
\hat{L}_z\right] \Psi,\label{GPE}
\end{eqnarray}
where $m$ is the atomic mass and $g=4 \pi \hbar^2 a_{\rm s}/m$
parameterizes the atomic interactions, with $a_{\rm s}$ being the
s-wave scattering length. The $\Omega \hat{L}_z$ term accounts for
frame rotation, where $\hat{L}_z=-i\hbar (x\frac{\partial}{\partial
y}-y\frac{\partial}{\partial x})$ is the {\it z}-component angular
momentum operator. We assume that the confining potential $V({\bf
r})$ is harmonic with the form,
\begin{eqnarray}
V({\bf
r})=\frac{1}{2}m\omega_r^2\left[(1-\epsilon)x^2+(1+\epsilon)y^2+\lambda^2
z^2\right],
\end{eqnarray}
where $\omega_r$ is the average trap frequency in the {\it x-y}
plane, $\epsilon$ is the trap ellipticity in the {\it x-y} plane,
and $\lambda=\omega_z/\omega_r$ is the trap ratio that determines the
axial trap strength.

In order to study the BSW solutions we employ a variational
technique.  This involves assuming a BSW ansatz and minimising its
energy to obtain the variational solutions.  This technique has
been employed for non-rotating BSWs and trapped attractive BECs, and
has been shown to have give very good agreement with the full
solution of the GPE. When the axial trapping is weak ($\lambda \ll
1$), we will employ the {\em sech
ansatz} for the BSW, given by,
\begin{eqnarray}
\Psi_{{\rm S}}&=&\sqrt{\frac{N}{2l_xl_yl_z
\pi}}e^{-\frac{x^2}{2l_x^2}} e^{-\frac{y^2}{2l_y^2}} {\rm sech}
\left(\frac{z}{l_z}\right)e^{i\alpha x y}, \label{P_Sech}
\end{eqnarray}
where $N$ is the atom number. Ignoring the $e^{i\alpha x y}$
term, this ansatz is identical to that used in Refs.
\cite{carr,salasnich,parker}, with the {\it sech} axial profile
appropriate because it is the form of the 1D soliton solution \cite{shabat}. The
term $e^{i\alpha x y}$ introduces a quadrupolar flow pattern to the
BEC, with $\alpha$ determining the amplitude. This flow pattern
preserves irrotationality and has been very successful in modelling
vortex-free repulsive condensates under rotation
\cite{Recati,Sinha,Rapid,PRL,madison2,corro}.

The total energy of the system is defined by,
\begin{eqnarray}
E=\int \left[\frac{\hbar^2}{2m}\left|\nabla \Psi\right|^2 +V({\bf
r})|\Psi|^2 +\frac{g}{2}|\Psi|^4 \right.\nonumber
\\
\left. +i\hbar\Omega\left(\Psi^{\star}x\frac{\partial\Psi}{\partial
y} -\Psi y\frac{\partial \Psi^{\star}}{\partial x}\right) \right]
d^3{\bf r}. \label{E_GPE}
\end{eqnarray}
Insertion of the ansatz into Eq. (\ref{E_GPE}) gives the
energy $E_{\rm S}$,
\begin{eqnarray}
\frac{E_{\rm S}}{N}&=&\frac{\hbar^2}{2m}\left[\frac{1}{2l_x^2}+\frac{1}{2l_y^2}+\frac{1}{3l_z^2}+\frac{l_x^2 \alpha^2}{2}+\frac{l_y^2 \alpha^2}{2}\right] \nonumber \\
&+& \frac{m \omega_r^2}{4}
\left[(1-\epsilon)l_x^2+(1+\epsilon)l_y^2+\frac{\lambda^2\pi^2}{6}l_z^2\right]\nonumber
\\&+&\frac{gN}{4\pi l_xl_yl_z} -\frac{\hbar
\Omega_z\alpha}{2}\left(l_x^2+l_y^2\right). \label{E_Sech}
\end{eqnarray}

We can reduce the number of variables in Eq. (\ref{E_Sech}) as
follows. Under the hydrodynamical interpretation, the
mean-field wavefunction can be expressed as $\Psi({\bf
r},t)=\sqrt{n({\bf r},t)}\exp[i\phi({\bf r},t)]$ where $n({\bf
r},t)$ and $\phi({\bf r},t)$ are the condensate density and phase,
respectively. Furthermore, ${\bf v}({\bf r},t)=(\hbar/m)\nabla \phi
({\bf r},t)$ is the ``fluid'' velocity. Inserting this into the GPE
and equating imaginary parts, one derives a continuity equation
given by,
\begin{eqnarray}
\frac{\partial n}{\partial t}+\nabla \cdot \left(n\left[{\bf v}-{\bf
\Omega}\times {\bf r}\right]\right)=0.
\end{eqnarray}
Our irrotational phase distribution $\phi({\bf r})=\alpha x y$
corresponds to a velocity distribution ${\bf v}({\bf r})=\hbar
\alpha(y{\bf {\hat i}}+x{\bf {\hat j}})/m$. Inserting this into the
continuity equation and setting $\partial n/\partial t=0$, we find
that stationary solutions satisfy,
\begin{eqnarray}
\alpha=\pm
\frac{m\Omega_z}{\hbar}\left(\frac{l_x^2-l_y^2}{l_x^2+l_y^2}\right).
\end{eqnarray}
Note that $\alpha$ can be positive or negative, resulting in two
``branches'' of solutions. For $\alpha>0$ the BSW is wider in the
{\it x}-direction than in the {\it y}-direction, and vice versa for
$\alpha<0$. We can thus eliminate $\alpha$ from Eq. (\ref{E_Sech}). For
simplicity we employ rescaled variables (in terms of the transverse
harmonic oscillator) $\gamma_x=l_x/a_r$, $\gamma_y=l_y/a_r$,
$\gamma_z=l_z/a_r$, ${\tilde \Omega}=\Omega/\omega_r$ and
$\varepsilon_{\rm S}=E_{\rm S}/(N\hbar \omega_r)$, where
$a_r=\sqrt{\hbar/m\omega_r}$ is the radial harmonic oscillator
length. Furthermore, we introduce the dimensionless interaction
parameter $k=N|a_{\rm s}|/a_r$. The ansatz energy becomes,
\begin{eqnarray}
\varepsilon_{{\rm
S}}&=&\frac{1}{4}\left[\frac{1}{\gamma_x^2}+\frac{1}{\gamma_y^2}+\frac{2}{3\gamma_z^2}\right.
\nonumber
\\ &+&\left.(1-\epsilon)\gamma_x^2+(1+\epsilon)\gamma_y^2+\frac{\lambda^2\pi^2}{6}\gamma_z^2\right]
\nonumber
\\
&-&\frac{k}{3\gamma_x\gamma_y\gamma_z}\pm
\frac{\tilde{\Omega}_z^2}{4}\left[\gamma_x^2-\gamma_y^2\right]\left[\frac{\gamma_x^2-\gamma_y^2}{\gamma_x^2+\gamma_y^2}-2\right].
\label{eps_Sech}
\end{eqnarray}

\begin{figure}[b]
\centering
\includegraphics[width=6.7cm]{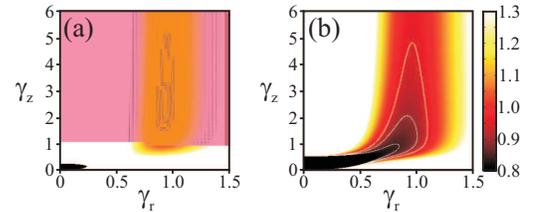}
\caption{Energy landscape of the non-rotating and non-elliptical
system for $\lambda=0$ according to Eq. (\ref{eps_Sech}). (a) Stable
regime $k=0.4<k_{\rm c}$ featuring a local energy minimum, i.e. the
BSW solution. (b) Unstable regime $k=0.8>k_{\rm c}$, where the whole
parameter space is unstable to collapse. White contours highlight
the shape of the landscapes. Since Eq. (\ref{eps_Sech}) is
cylindrically symmetric in this case, we introduce a radial
lengthscale $\gamma_r=\sqrt{(\gamma_x^2+\gamma_y^2)/2}$.}
\end{figure}

Recall that Eq. (\ref{eps_Sech}) is valid for $\lambda \ll 1$. Under
tight axial trapping $\lambda\gg0$, this direction is dominated by
the trap rather than the interactions and it is more appropriate to
consider a {\em gaussian ansatz},
\begin{equation}
\Psi_{{\rm G}}=\sqrt{\frac{N}{l_xl_yl_z
\pi^{3/2}}}e^{-\frac{x^2}{2l_x^2}}e^{-\frac{y^2}{2l_y^2}}e^{-\frac{z^2}{2l_z^2}}e^{i\alpha
x y}.
\end{equation}

The rescaled energy for the gaussian ansatz is then,
\begin{eqnarray}
\varepsilon_{{\rm
G}}&=&\frac{1}{4}\left[\frac{1}{\gamma_x^2}+\frac{1}{\gamma_y^2}+\frac{1}{\gamma_z^2}\right.
\nonumber
\\ &+&\left.(1-\epsilon)\gamma_x^2+(1+\epsilon)\gamma_y^2+\lambda^2\gamma_z^2\right]
\nonumber
\\
&-&\frac{k}{\sqrt{2\pi}\gamma_x\gamma_y\gamma_z}\pm
\frac{\tilde{\Omega}_z^2}{4}\left[\gamma_x^2-\gamma_y^2\right]\left[\frac{\gamma_x^2-\gamma_y^2}{\gamma_x^2+\gamma_y^2}-2\right].
\label{eps_Gauss}
\end{eqnarray}
Note that, for the regimes of interest, the sech and gaussian ansatz
give similar results, typically differing by less than $10\%$
\cite{parker}. We can now map out the 3D energy landscapes of the
rotating BSWs as a function of the lengthscales $\gamma_x$,
$\gamma_y$ and $\gamma_z$. Variational BSW solutions exist where
there is a local energy minimum in the landscape, and are obtained
by a numerical search algorithm. The local energy minimum has widths
$\gamma^0_x$, $\gamma^0_y$ and $\gamma^0_z$, energy $\varepsilon_0$,
and quadrupolar flow amplitude $\alpha_0$.

We first revisit the  $\lambda=0$ BSW solutions in the
absence of rotation and ellipticity, as studied
previously using the $\alpha=0$ limits of Eq. (\ref{eps_Sech})
\cite{carr,parker} and Eq. (\ref{eps_Gauss}) \cite{perez_gaussian}.
Here the energy landscape is
cylindrically symmetric ($\epsilon=0$) and so we introduce a radial lengthscale
$\gamma_r=\sqrt{(\gamma_x^2+\gamma_y^2)/2}$. A typical energy
landscape, according to Eq. (\ref{eps_Sech}), for a stable BSW
solution is presented in Fig. 1(a), corresponding to $k=0.4$. At the
origin the interaction term in Eq. (\ref{eps_Sech}) diverges to
negative values and is a region of collapse of the BSW. However,
there exists a local energy minimum which represents
the self-trapped BSW solution. A typical unstable energy landscape
is shown in Fig. 1(b) for $k=0.8$. No local energy minimum exists,
and the whole parameter space is unstable to collapse. From this the
critical interaction strength for collapse is determined to be
$k_{\rm c}=0.76$, in good agreement with
full solution of the GPE which gives $k_{\rm c}\approx 0.68$ \cite{carr,parker}.

\begin{figure}[t]
\centering
\includegraphics[width=5.8cm,clip=true]{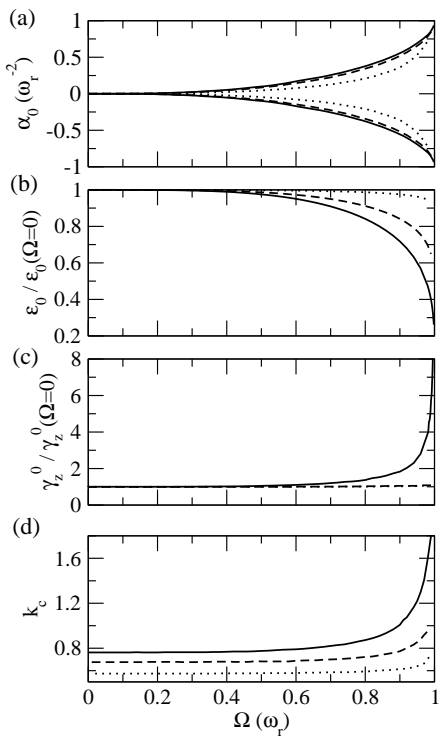}
\caption{Properties of $\epsilon=0$ BSW solutions as a function of
$\Omega$ for $\lambda=0$ (solid line) and the axially-trapped cases
of $\lambda=1$ (dashed lines) and $\lambda=5$ (dotted lines). (a)
Quadrupolar flow amplitude $\alpha_0$ for interaction parameter
$k=0.4$. (b) BSW energy $\varepsilon_0$ for $k=0.4$. (c) Axial
lengthscale $\gamma^0_z$ for $k=0.4$. (d) Critical interaction
parameter for collapse $k_c$. Note that for $\lambda=0$ ($\lambda>0$) we employ
the sech (gaussian) ansatz.}
\end{figure}
\begin{figure}
\centering
\includegraphics[width=5.7cm,clip=true]{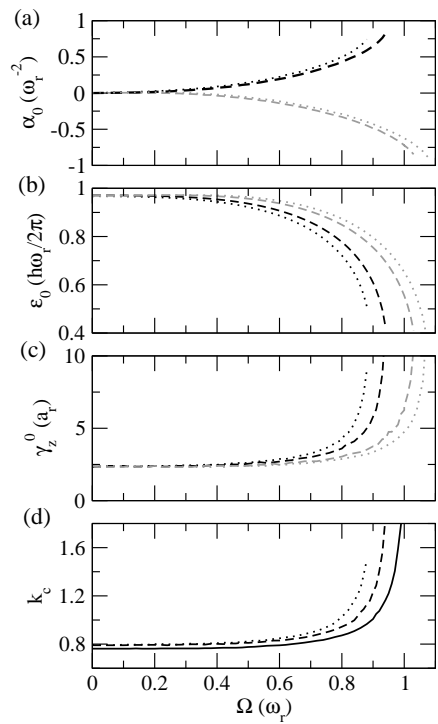}
\caption{Properties of the rotating $\lambda=0$ BSWs for
elliptical traps of $\epsilon=0.1$ (dashed lines) and $0.2$
(dotted lines). Black (grey) lines indicate upper (lower) branch
solutions. (a) Quadrupolar flow amplitude $\alpha_0$ for $k=0.4$.
(b) BSW energies $\varepsilon_0$, rescaled by their
non-rotating values of $\varepsilon_0(\Omega=0)=0.972$ for
$\lambda=0$, $1.315$ for $\lambda=1$ and $3.01$ for $\lambda=5$. (c)
Axial lengthscales $\gamma^0_z$, rescaled by $\gamma^0_z
(\Omega=0)=2.36$ for $\lambda=0$, $0.902$ for $\lambda=1$ and
$0.416$ for $\lambda=5$. (d) Critical interaction parameter for
collapse $k_c$ of the upper branch (lowest energy) solutions.  }
\end{figure}

We will now consider the effect of rotation.  For simplicity we
first assume $\epsilon=0$. Figure 2(a)-(c) shows how the key
parameters vary as rotation is introduced for a fixed interaction
parameter $k=0.4$. For $\Omega
> 0$ the symmetry between $\gamma_x$ and $\gamma_y$ is broken and
the solutions have non-zero $\alpha_0$ (Fig. 2(a)). We see the
formation of two branches of $\alpha_0$.  Due to the trap symmetry in the {\it x-y}  plane, the branches are symmetric about the
$\alpha_0=0$ axis, with the upper branch being elongated in the {\it
x}-direction and the lower branch being elongated in the {\it
y}-direction. As $\Omega$ increases, so too does the magnitude of
$\alpha_0$, implying a spreading of the BSW in the {\it x-y} plane.
This is because of the growth of an outward centrifugal force. As
$\Omega$ approaches $\omega_r$, $\alpha_0$ diverges to $\pm \infty$.
This is because, at $\Omega=\omega_r$, the centrifugal force exactly
balances the trapping potential, and the BEC is untrapped in the
{\it x-y} plane. Since the BEC centre-of-mass becomes dynamically
unstable, this is termed the {\em centre-of-mass instability}
\cite{rosenbusch}. The BSW energy (Fig. 2(b)) decreases towards zero
as $\Omega \rightarrow \omega_r$ as a result of the reduced density.
The axial lengthscale (Fig. 2(c)) grows with $\Omega$ since the
radial spreading dilutes the interaction strength and forms a
less tightly bound BSW.

We have isolated the critical interaction parameter for collapse
$k_c$ as a function of $\Omega$ with the results shown in Fig. 2(d).
The most striking feature is that $k_{\rm c}$ dramatically increases
as $\Omega_z$ approaches  $\omega_r$. This is directly associated with
the radial spreading and reduced density of the rotating solutions.
Specifically, for $\Omega/\omega_r =0.9$, $k_c$ is approximately $50\%$ larger
than its non-rotating value while for $\Omega/\omega_r =0.97$, $k_c$ is approximately twice as large.

In Fig. 2 we also consider the presence of significant axial trapping
$\lambda=1$ (dashed line) and $5$ (dotted line), for which we employ the gaussian ansatz.  We see
similar qualitative behaviour to the $\lambda=0$ case: a
divergent growth of $\alpha_0$ (Fig. 2(a)) and decrease in
$\varepsilon_0$ (Fig. 2(b)).  However, the magnitudes are
consistently less than the corresponding $\lambda=0$ results.  The
axial lengthscales (Fig. 2(c)) show little variation with $\Omega$
since this is now dominated by the external axial trapping.  The
critical point for collapse $k_{\rm c}$ grows with $\Omega$, but at
a slower rate than for $\lambda=0$.  Note that the presence of axial
trapping reduces $k_{\rm c}$, as observed elsewhere
\cite{parker,gammal}.

The $\epsilon=0$ limit is somewhat unphysical since no torque is
actually applied to the BEC. We now consider the more realistic case
of finite trap ellipticity. The results for $\lambda=0$ BSWs under
$\epsilon=0.1$ and $0.2$ are presented in Fig. 3.  The finite
ellipticity breaks the symmetry in the {\it x-y} plane and therefore
in the branches of $\alpha_0$. The upper branch solutions are
elongated in the {\it x}-direction, which has trap frequency
$\omega_x=\sqrt{(1-\epsilon)}\omega_r$.  The centre-of-mass
instability then occurs when $\Omega=\sqrt{(1-\epsilon)}\omega_r$,
which is why the divergence in $\alpha_0$ shifts to lower $\Omega$.
Conversely, the lower branch solutions diverge at
$\Omega=\sqrt{1+\epsilon}\omega_r$ and so become shifted towards
larger $\Omega$.

Consideration of the energy (Fig. 3(b)) shows that
the upper branch solutions have lower energy.  That is, it is lower
energy for the condensate to be elongated in {\it x} than {\it y},
since the trap is weaker in this direction.  Consequently, we expect
that only the upper branch solutions would ever be observed.  Apart
from the shift in the asymptotes introduced by the finite
ellipticity, the behaviour of $\alpha_0$, $\epsilon_0$ and
$\lambda^0_z$ is qualitatively and quantitatively similar to the
$\epsilon=0$ case. In Fig. 2(d) we plot the critical interaction
parameter for collapse $k_{\rm c}$ of the upper branch solutions.
Again, we see similar behaviour to the $\epsilon=0$ case, with a
dramatic in increase in $k_{\rm c}$ as $\Omega$ approaches
$\sqrt{1-\epsilon}\omega_r$.

In this work we have employed a variational technique to study
bright solitary matter-wave solutions under rotation in elliptical
traps. This is made possible by incorporating an irrotational
quadrupolar flow pattern into the variational ansatz. Importantly,
the BSW becomes more stable to the collapse instability when under
rotation.  This is most pronounced when the rotation frequency $\Omega$ is slightly less than the minimum effective trap frequency $\sqrt{1-\epsilon}\omega_r$, e.g., for
$\Omega=0.9\sqrt{1-\epsilon}\omega_r$, the critical interaction
parameter for collapse $k_{\rm c}$ increases by approximately $50\%$ over
the non-rotating value. This implies that the BSW can support the
same increase in number of atoms, before collapse.

We acknowledge funding from the ARC and the
Canadian Commonwealth Scholarship Program (NGP).

\end{document}